\documentclass[11pt]{article}
\usepackage{deauthor}
\usepackage{times, graphicx}
\usepackage{natbib}
\usepackage{booktabs}
\usepackage{hyperref}
\graphicspath{{xiekun/}{figs/}}

\title{A Survey on False Information Detection: From A Perspective of Propagation on Social Networks}

\author{Kun Xie \\[1ex] \small{The Chinese University of Hong Kong}\\[1ex] \small{\texttt{xiekun@se.cuhk.edu.hk}} \and Sibo Wang \\[1ex] \small{The Chinese University of Hong Kong}\\[1ex] \small{\texttt{swang@se.cuhk.edu.hk}}}

\begin{document}

\maketitle

\begin{abstract}
The proliferation of false information in the digital age has become a pressing concern, necessitating the development of effective and robust detection methods. This paper offers a comprehensive review of existing false information detection techniques, approached from a novel perspective that emphasizes the propagation characteristics of misinformation. We introduce a new taxonomy that categorizes these methods into homogeneous and heterogeneous propagation-based approaches, providing a deeper understanding of the varying scopes and complexities involved in information dissemination. For each category, we present a formal problem formulation, review commonly used datasets, and summarize state-of-the-art methods. Additionally, we identify several promising directions for future research, including the creation of a unified benchmark suite, exploration of diverse information modalities, and development of innovative rumor debunking tasks. By systematically organizing the vast array of current techniques, this work offers a clear overview of the research landscape, aiding researchers and practitioners in navigating this complex field and inspiring further advancements.
\end{abstract}

\section{Introduction}
In the digital era, social media platforms facilitate both the rapid dissemination of information and the spread of false content, such as rumors and fake news. These can have severe consequences, including misleading the public~\cite{lazer2018science}, inciting panic~\cite{vosoughi2018spread}, and influencing elections~\cite{bakdash2018future, jin2017detection}. Detecting and debunking false information early is crucial to mitigate its harmful impacts on society. 
The way information spreads through social networks can reveal its veracity~\cite{ma2018rvnn, bian2020bigcn}. False information often exhibits distinct propagation patterns, such as rapid spread and specific dissemination strategies. By analyzing these propagation patterns and characteristics, researchers can identify false content more effectively and obtain deeper insights into how misinformation spreads.

This study seeks to offer a comprehensive understanding of propagation-based false information detection, addressing a gap left by existing surveys. Previous works~\cite{hussain2025fake, taylor2024misinformation, shu2017fake, guo2020future} primarily focus on text content-based rumor detection using general deep learning techniques. 
Other studies~\cite{li2025survey, mostafa2024modality} emphasize the fusion of multiple modalities, integrating text and images to enhance detection capabilities. 
Additionally, some research targets specific aspects of false information detection. For example, \citep{kwao2025survey} concentrates on early detection, while \citep{battista2025survey} examines detection within particular populations. Despite these contributions, the propagation characteristics of false information on social networks remain underexplored. This perspective offers valuable insights that extend beyond traditional content analysis, highlighting the dynamic nature of misinformation spread and its implications for detection strategies.
In recent years, several surveys~\cite{mahdi2024survey, gong2023fake, lakzaei2024disinformation, phan2023fake} have summarized methods for false information detection utilizing Graph Neural Networks (GNNs)~\cite{kipf2016semi, hamilton2017inductive}. However, these surveys predominantly concentrate on GNN techniques, lacking a comprehensive review and analysis from the perspective of propagation information. While propagation information can indeed be modeled using graph topology, it is important to note that GNNs are not the sole technology employed in this domain. Architectures such as Transformers~\cite{beltagy2020longformer, vaswani2017attention, devlin2019bert} and RNNs~\cite{cho2014properties, ma2018rvnn} have also been extensively studied and applied. Consequently, focusing solely on GNN-based methods does not provide a complete overview of propagation-based approaches. Our work seeks to address this gap by offering a more holistic understanding of false information detection through the lens of propagation dynamics.

Specifically, we begin by reviewing the definition of false information and introducing the concept of information propagation on social networks. We categorize propagation based on its homogeneous or heterogeneous nature. Homogeneous propagation involves the dissemination of a source post through user interactions such as comments and retweets, forming a single-type propagation graph where all nodes represent posts. This type of propagation provides additional perspectives and debates that aid in assessing the veracity of the original news. In contrast, heterogeneous propagation encompasses a broader social context, including user metadata, semantic information, and cross-platform discussions. This rich context forms a heterogeneous social context graph with multiple node types, offering a comprehensive view of news propagation and enhancing veracity understanding. These two categories reflect different scopes and complexities of information propagation, each presenting unique insights and challenges for rumor detection.

\begin{figure}
    \centering
    \includegraphics[width=0.75\linewidth]{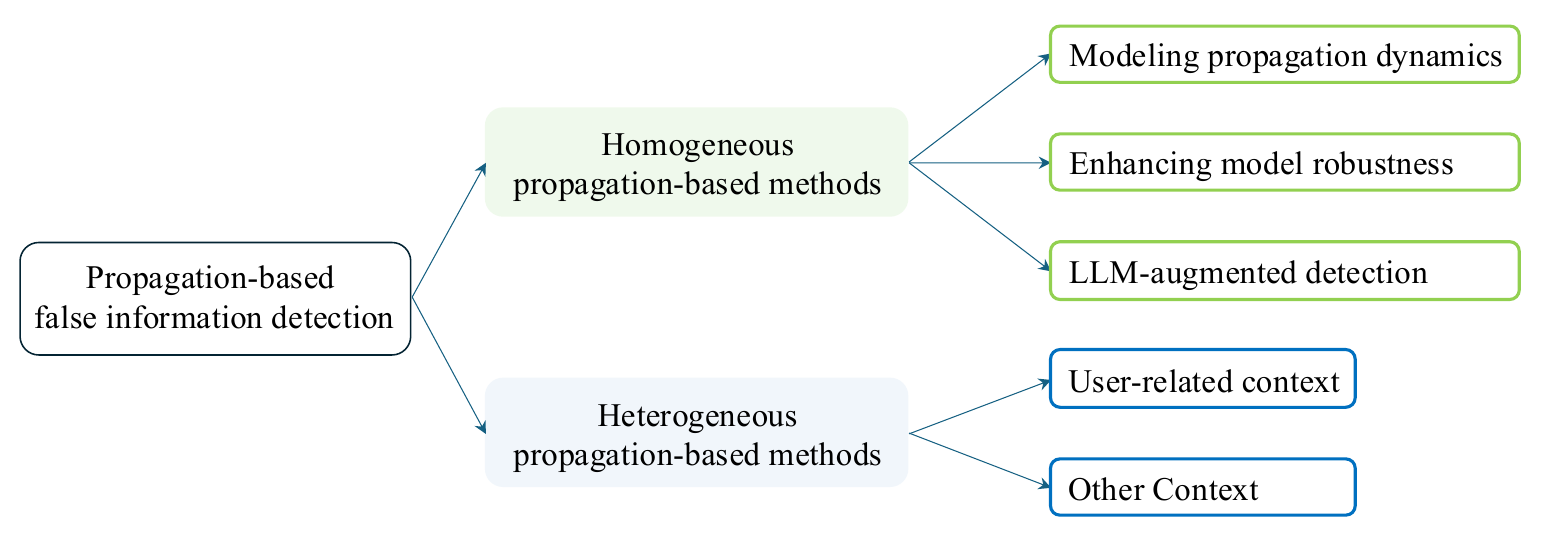}
    \caption{Taxonomy of propagation-based false information detection methods. }
    \label{fig:taxonomy}
\end{figure}

Based on this categorization, we have reviewed false information detection methods that utilize both homogeneous and heterogeneous propagation. For each category, we first provide a formal problem formulation to help readers better understand the task. We then review the commonly used datasets and summarize existing methods. For methods based on homogeneous propagation, we categorize them from the perspective of propagation into three subcategories: 1) methods that model the dynamics of propagation, 2) methods that enhance robustness in the face of complex and noisy real-world propagation, and 3) methods that utilize Large Language Models (LLMs) to aid in understanding propagation. For methods based on heterogeneous propagation, we categorize them based on the types of social context they incorporate into two subcategories: 1) user-related context and 2) other context.
Finally, we propose several feasible future research directions, including the development of an off-the-shelf unified benchmark suite and exploring alternative approaches to mitigate the impact of false information, to promote the advancement of this field. Figure~\ref{fig:taxonomy} illustrates the taxonomy proposed in this paper.

The contributions of this work can be summarized as follows:
\begin{itemize}
\item \textit{New Perspective}. This study reviews and synthesizes existing false information detection methods from the perspective of propagation, categorizing them into homogeneous and heterogeneous types. This approach provides insights into the varying scopes and complexities involved in information dissemination.

\item \textit{New Taxonomy}. We propose a novel taxonomy for propagation-based false information detection methods, initially dividing them based on their use of homogeneous or heterogeneous propagation. This taxonomy aids in systematically organizing the extensive array of existing methods.

\item \textit{Comprehensive Review and Summary}. This work offers a thorough and detailed review of existing methods for false information detection. For each propagation category, we present a formal problem formulation, review commonly used datasets, and summarize state-of-the-art methods. This comprehensive overview provides readers with a holistic understanding of the current research landscape in this field.
\end{itemize}

This paper is organized as follows: Section~\ref{sec:definition} introduces the concept of false information and commonly used terms in the field. Section~\ref{sec:prop_cate} explores the characteristics and distinctions between homogeneous and heterogeneous propagation. Sections~\ref{sec:homo_prop} and ~\ref{sec:hetero_prop} provide comprehensive summaries of existing false information detection methods based on homogeneous and heterogeneous propagation, respectively. Each section includes a formal problem formulation, a review of commonly used datasets, and an in-depth analysis of state-of-the-art methods. Finally, Section~\ref{sec:future} discusses potential future research directions, highlighting opportunities for further advancement in this area.

\section{False Information on Social Media}
\label{sec:definition}

To comprehend false information on the web and social media, \citet{kumar2018false} systematically introduces this concept by categorizing it according to intent and knowledge content. Based on intent, false information can be classified as \textit{misinformation}, which is created without the intention to mislead, and \textit{disinformation}, which is deliberately produced to deceive and mislead readers~\cite{fallis2014functional}. From the perspective of knowledge content, false information can be divided into \textit{opinion-based}, where a unique ground truth does not exist (as in the case of product reviews on e-commerce platforms), and \textit{fact-based}, which involves falsehoods about entities that have a unique, verifiable ground truth~\cite{thomas1986statements}.

In the field of false information detection, two terms are frequently encountered. The first term is \textit{fake news}~\cite{shu2017fake}, which refers to deliberately false or verifiably inaccurate news content, falling within the category of disinformation under the aforementioned taxonomy of false information. The second term is \textit{rumor}, defined as unverified and instrumentally relevant information statements in circulation~\cite{dataset_weibo, guo2020future}. Unlike fake news, rumors may ultimately be proven true or false~\cite{zubiaga2018detection}. Despite their conceptual differences, these two terms are often used interchangeably in the context of social media false information detection, with their subtle distinctions being largely overlooked in previous studies~\cite{gong2023fake}. Following this established practice in the literature, this survey reviews previous false information detection studies targeting both fake news and rumors.

False information can be disseminated through various types of media on social networks. In existing datasets, the common carriers include \textit{micro-posts} on platforms such as Twitter\footnote{\url{https://x.com/}} and sina Weibo\footnote{\url{https://m.weibo.cn/}}, as well as published fake news \textit{articles}. These carriers exhibit distinct characteristics: micro-posts are typically brief and colloquial, while news articles are longer and more structured, with separate components such as titles and content. While these structural differences are noteworthy, they are not the focus of our discussion. Instead, this paper emphasizes how false information propagates through social networks, regardless of its carrier type. In the next section, we present two primary approaches that existing works have employed to capture these propagation patterns.

\section{Information Propagation on Social Networks}
\label{sec:prop_cate}
The propagation of false information on social networks can be categorized into two distinct types based on the nature of propagation:

\begin{itemize}
    \item \textbf{Homogeneous propagation} refers to the dissemination of a source post on social media, where other users interact with it through comments or retweets. These interactions, which may include expressing opinions on the source news or engaging in discussions with other comments, provide additional perspectives, supplementary information, and debates regarding the original news. The textual content of these retweets and comments offers further evidence for classifying the veracity of the original news. Moreover, the reply relationships between these discussions form a graph topology, linking the relevant texts together and further facilitating the prediction of veracity. We term this homogeneous propagation because it involves only a single type of propagation, forming a homogeneous propagation graph where all nodes represent posts on social media.
    
    \item \textbf{Heterogeneous propagation}, in contrast, encompasses the spread of information across broader social networks. Specifically, in addition to direct interactions with the original news, user metadata such as profiles and friendships, semantic information of the original news like keywords and domains, and related discussions of the news on other social platforms are all captured as social context. This rich social context provides a comprehensive view of the news propagation and discussion, enhancing the understanding of the news's veracity. We refer to this as heterogeneous propagation because it involves multiple types of propagation forms. This heterogeneous propagation can be modeled with a heterogeneous social context graph, where nodes represent not just social media posts but also news articles, cross-platform discussions, and user accounts, while the edges connect different types of nodes, reflecting explicit and implicit information flow on social networks.
\end{itemize}

\begin{figure}
    \centering
    \includegraphics[width=0.65\linewidth]{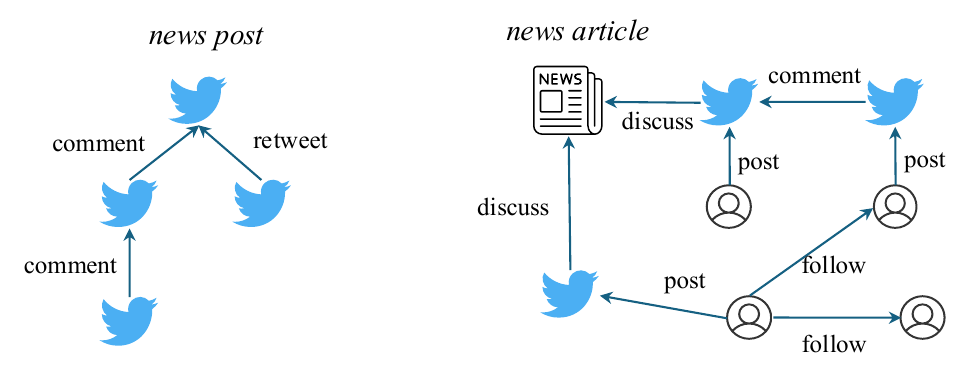}
    \caption{Illustration of homogeneous propagation (left) and heterogeneous propagation (right). }
    \label{fig:two-prop}
\end{figure}

These two categories reflect the different scopes and complexities of information propagation on social networks, each offering unique insights and challenges for rumor detection. Figure~\ref{fig:two-prop} illustrates these two types of propagation. The left sub-figure displays the homogeneous propagation of a post on social media, with commenting and retweeting interaction posts forming the homogeneous propagation tree. Conversely, the right sub-figure demonstrates the propagation of a news article across complex social networks, where the news article is connected to related discussion posts on other social media platforms, and the user profiles and user friendship network are also collected, together forming a heterogeneous social context graph.

Based on these two distinct propagation mechanisms, we proceed to provide a comprehensive analysis of each approach. In Section~\ref{sec:homo_prop}, we present a formal problem formulation for homogeneous propagation-based rumor detection, followed by an overview of commonly used datasets and a systematic review of existing methodologies. Similarly, Section~\ref{sec:hetero_prop} examines heterogeneous propagation-based false information detection, detailing its mathematical formulation, popular datasets, and a thorough survey of current research efforts.

\section{Methods Based on Homogeneous Propagation}
\label{sec:homo_prop}
\subsection{Problem Formulation}
\label{sec:prob1}
The task under this category is to classify a given post as either a rumor or a non-rumor, based on interactions occurring on the same social media platform, include commenting and retweeting the source post.

Formally, the propagation topology of the given source post can be modeled by a homogeneous graph, $G_p=(V_p, E_p, X_p)$, where the subscript $p$ denotes the index of the source post. Specifically, the graph $G_p$ is acyclic and tree-structured, as each comment or retweet is directed towards a single parent post. The node set $V_p$ comprises all nodes in the graph $G_p$, with $v_0$ as the root representing the initial source post, and each $v_i$ for $i \in \{1, \cdots, \vert V_p \vert -1\}$ representing a comment or retweet post. The edge set $E_p$ includes all edges in the graph $G_p$, where each edge $(v_i, v_j)$ signifies the comment or retweet relationship between two posts, i.e., post $v_j$ comments on or retweets post $v_i$. The set $X_p$ contains the text content of all posts, where $i$-th row of $X_p$ denotes the text attributes of the corresponding node $v_i$. The source post, along with its direct propagation graph, is referred to as an event~\cite{bian2020bigcn}, and the dynamics of commenting and retweeting are sometimes called conversation threads~\cite{ma2018rvnn}.

The objective is to predict the veracity of the given post using both the text content and the direct propagation information $G_p$. The definition of rumor veracity varies across datasets. For instance, in the Twitter15~\cite{dataset_twitter15_twitter16} and Twitter16~\cite{dataset_twitter15_twitter16} datasets, posts are categorized into \textit{1)} non-rumors, \textit{2)} false rumors, \textit{3)} true rumors, and \textit{4)} unverified rumors. Conversely, in other datasets like PHEME5~\cite{dataset_pheme5_1, dataset_pheme5_2} and PHEME9~\cite{dataset_pheme9}, posts are classified as \textit{1)} rumors and \textit{2)} non-rumors.

In a typical supervised learning framework, a function $f:\mathcal{G}\rightarrow\mathcal{Y}$ is learned to predict veracity labels for posts with propagation graphs, where $\mathcal{G}$ represents the space of events, and $\mathcal{Y}$ denotes the space of veracity labels. The function $f$ is trained using a labeled training set $\mathcal{G}_{train}$, where the ground-truth veracity label $y_p$ is available for each $G_p\in\mathcal{G}_{train}$. The learned function $f$ is then applied to classify unlabeled events in the test set $\mathcal{G}_{test}$. Most existing propagation-based rumor detection studies follow this supervised learning approach, although some employ self-supervised learning or transfer learning paradigms, as will be discussed later.

\begin{figure}
    \centering
    \includegraphics[width=0.9\linewidth]{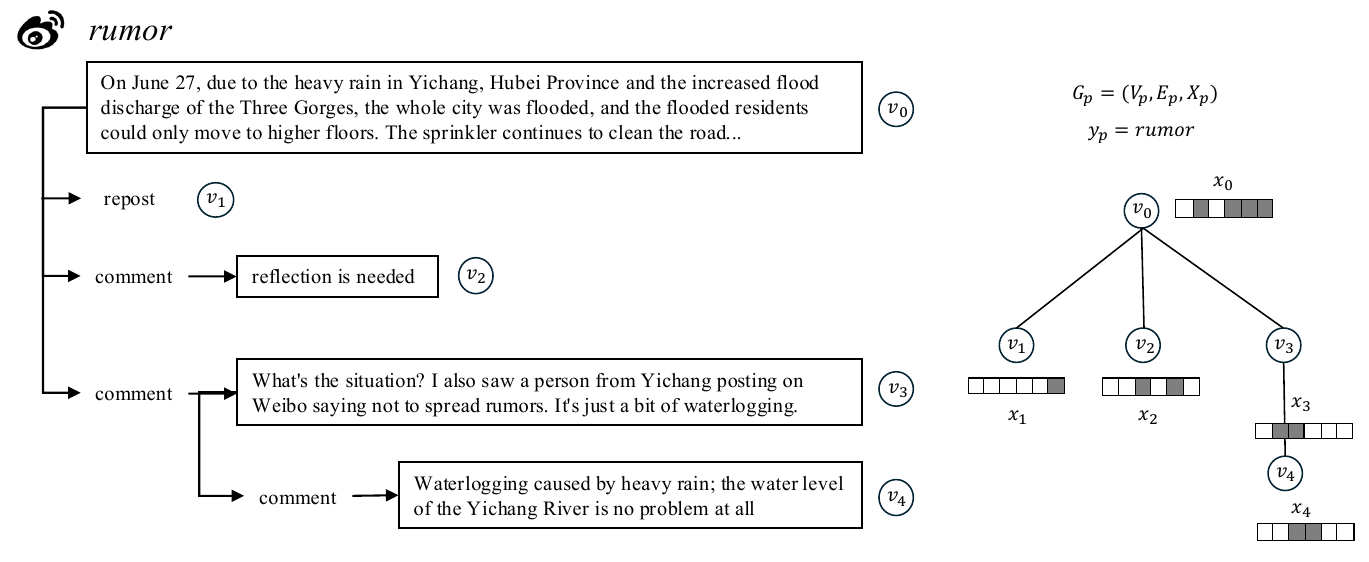}
    \caption{An example event from Weibo22~\cite{dataset_weibo22} containing false information related to a flood disaster, and the corresponding graph $G_p$ modeling the propagation threads.  }
    \label{fig:example1}
\end{figure}

Figure~\ref{fig:example1} illustrates an example event from the Weibo22 dataset~\cite{dataset_weibo22}, showcasing how false information about a flood spread on the Sina Weibo platform. The post claims that Yichang City in China experienced severe flooding, allegedly due to heavy rainfall and increased water discharge from the Three Gorges Dam. As this information circulated on the platform, users engaged through reposts and comments, reflecting the propagation threads and patterns of the source post. Notably, the comment posts contributed valuable, diverse perspectives and discussions, playing a crucial role in collectively verifying the authenticity of the source information. For instance, local users reported only minor waterlogging in subsequent comments and called for thorough fact-checking. By abstracting both the source post and its associated comments/reposts as nodes, the entire propagation process can be represented as a homogeneous graph, as depicted on the right side of the figure.

\begin{table}[t]
\caption{Statistics of commonly used datasets, PHEME5~\cite{dataset_pheme5_1, dataset_pheme5_2}, PHEME9~\cite{dataset_pheme9}, Weibo-COVID19~\cite{dataset_weibo_covid19_twitter_covid19}, and Twitter-COVID19~\cite{dataset_weibo_covid19_twitter_covid19}, for direct propagation-based false information detection. }
\label{tab:homo-stat1}
\vspace{2ex}
\centering
\small
\begin{tabular}{lrrrrrrr}
\toprule
 & \multicolumn{1}{l}{PHEME5} & \multicolumn{1}{l}{PHEME9} & \multicolumn{1}{l}{\begin{tabular}[c]{@{}l@{}}Weibo-\\ COVID19\end{tabular}} & \multicolumn{1}{l}{\begin{tabular}[c]{@{}l@{}}Twitter-\\ COVID19\end{tabular}} & \multicolumn{1}{l}{Weibo} & \multicolumn{1}{l}{Weibo21} & \multicolumn{1}{l}{Weibo22} \\
 \midrule
\# Source posts & 5,802 & 6,425 & 399 & 400 & 4,664 & 7,567 & 4,174 \\
\begin{tabular}[c]{@{}l@{}}\# Comment/retweet \\ posts\end{tabular} & 103,212 & 105,354 & 26,687 & 406,185 & 3,805,656 & 184,775 & 961,962 \\
\begin{tabular}[c]{@{}l@{}}Avg. \# comment \\ per event\end{tabular} & 18 & 16 & 67 & 1015 & 816 & 24 & 230 \\
\midrule
\# Non-rumor & 3,830 & 4,023 & 146 & 148 & 2,351 & 3,605 & 2,087 \\
\# Rumor & 1,972 & 2,402 & 253 & 252 & 2,313 & 3,962 & 2,087 \\
\bottomrule
\end{tabular}
\end{table}

\subsection{Datasets}
\label{sec:dataset1}
We have compiled a collection of datasets commonly utilized in existing studies on homogeneous propagation-based detection of false information. Detailed descriptions of these datasets are provided below:

\begin{itemize}
\item The \textbf{PHEME5}~\cite{dataset_pheme5_1, dataset_pheme5_2} dataset was constructed by harvesting tweets from Twitter that pertain to newsworthy events with the potential to incite and propagate rumors. This dataset collected tweets based on five such events: \textit{the Ferguson unrest, the Ottawa shooting, the Sydney crisis, the Charlie Hebdo shooting}, and \textit{the Germanwings flight crash}. The collected tweets were then identified as rumors and non-rumors by journalists based on the source tweet contents and their associated comment threads.

\item The \textbf{PHEME9}~\cite{dataset_pheme9} dataset includes posts and propagation graphs derived from comments and reposts on Twitter. It encompasses tweets related to nine events, including breaking news likely to generate multiple rumors and known rumor events. Specifically, the dataset covers five sudden news events: \textit{the Ferguson unrest, the Ottawa shooting, the Sydney crisis, the Charlie Hebdo shooting}, and \textit{the Germanwings flight crash}. Additionally, it includes four specific rumors: \textit{the Prince concert in Toronto, the Gurlitt collection, Putin's disappearance}, and \textit{Michael Essien contracted Ebola}. Similar to PHEME5, journalists annotated the source posts for the collected tweets as either rumors or non-rumors, along with their propagation threads.

\item The \textbf{Weibo}~\cite{dataset_weibo} dataset was constructed based on Sina Weibo, a large Chinese microblog platform. The dataset comprises a set of known rumors sourced from the Sina community management center\footnote{\url{https://service.account.weibo.com/}}, documenting various instances of false information. An equivalent number of non-rumor events were gathered by randomly crawling posts not reported as rumors.

\item The \textbf{Weibo21}~\cite{dataset_weibo21} dataset is a multi-domain fake news dataset curated based on the Sina Weibo platform. It includes rumor events collected from reports by the Weibo Community Management Center, while non-rumor events were collected randomly and further verified by NewsVerify\footnote{\url{https://www.newsverify.com/}}. All news items were annotated across nine domains: \textit{Science, Military, Education, Disasters, Politics, Health, Finance, Entertainment,} and \textit{Society} using crowd-sourcing.

\item The \textbf{Weibo22}~\cite{dataset_weibo22} dataset was designed to address the issue of incomplete and untraceable posts in the previous Weibo and Weibo21 datasets. Specifically, Weibo22 provides original comment text and comprehensive author profile details. Events in the dataset are categorized as rumors or non-rumors based on information from the Weibo Community Management Center and the China Internet Joint Rumor Debunking Platform\footnote{\url{https://www.piyao.org.cn/}}.Over half of the posts in Weibo22 relate to the COVID-19 pandemic.

\item The \textbf{Weibo-COVID19} and \textbf{Twitter-COVID19}~\cite{dataset_weibo_covid19_twitter_covid19} datasets were developed focusing on the COVID-19 pandemic. The Weibo-COVID19 dataset comprises rumor posts on Sina Weibo related to COVID-19, annotated by the Sina community management center, and non-rumor posts collected randomly. The Twitter-COVID19 dataset extends text-only claims from a COVID-19 rumor dataset~\cite{kar2021no} by incorporating propagation threads from Twitter. These datasets include micro posts in Chinese and English, respectively.

\item The \textbf{Twitter15} and \textbf{Twitter16}~\cite{dataset_twitter15_twitter16} datasets extend previous text-only datasets by incorporating propagation trees. Twitter15 was constructed based on the framework proposed in~\cite{liu2015real}, and Twitter16 was an extension of the dataset in~\cite{dataset_weibo}. Furthermore, the original binary labels (rumors and non-rumors) were refined to include four categories: non-rumors, false rumors, true rumors, and unverified rumors, in accordance with the veracity tags on articles from rumor debunking websites\footnote{\url{snopes.com} and \url{Emergent.info}}.
\end{itemize}

Tables~\ref{tab:homo-stat1} and \ref{tab:homo-stat2} present common statistics for nine popular datasets. Specifically, Table~\ref{tab:homo-stat1} provides statistics for seven datasets where news posts are classified into rumors and non-rumors, while Table~\ref{tab:homo-stat2} displays statistics for two datasets with news posts classified into four categories: non-rumors, false rumors, true rumors, and unverified rumors. It is important to note that the statistics of these datasets may vary over time due to the dynamic nature of social networks, where users frequently delete posts or deactivate their accounts. Consequently, this can lead to inconsistencies in dataset statistics across different studies.

\begin{table}[t]
\caption{Statistics of commonly used datasets, Twitter15~\cite{dataset_twitter15_twitter16} and Twitter16~\cite{dataset_twitter15_twitter16}, for direct propagation-based false information detection.}
\label{tab:homo-stat2}
\begin{center}
\begin{tabular}{lrr}
\toprule
 & \multicolumn{1}{l}{Twitter15} & \multicolumn{1}{l}{Twitter16} \\ 
 \midrule
\# Source posts & 1,490 & 818 \\
\# Comment/retweet posts & 331,612 & 204,820 \\
Avg. \# comments per event & 223 & 251 \\
\midrule
\# Non-rumor & 374 & 205 \\
\# False rumor & 370 & 205 \\
\# True rumor & 372 & 205 \\
\# Unverified rumor & 374 & 203 \\
\bottomrule
\end{tabular}
\end{center}
\end{table}

\subsection{Method Review}
\label{sec:methods1}
In this section, we review related works that detect false information on social media based on direct propagation threads. We categorize existing methods into three categories: 1) methods that focus on modeling the dynamics of propagation, 2) methods that focus on improving the robustness of propagation-based detection, and 3) methods that utilize large language models (LLMs) to augment propagation. We elaborate on the related methods from these aspects in the following sections.

\subsubsection{Dynamics of Propagation}
RvNN~\cite{ma2018rvnn} and BiGCN~\cite{bian2020bigcn} model the dynamics of propagation threads through two distinct directions: the bottom-up direction, which models rumor dispersion, and the top-down direction, which models rumor propagation. In the bottom-up tree, responsive nodes always point to the nodes they are responding to, resembling a citation network where a response acts like a reference. Conversely, the top-down structure aligns with the natural flow of information, where a link from node $v_i$ to $v_j$ indicates that $v_j$ reads the information from $v_i$ and responds to it. This setup simulates how information cascades from the source tweet to all its receivers. Based on these two directed tree structures, RvNN and BiGCN utilize recursive neural models~\cite{cho2014properties} and graph neural networks~\cite{kipf2016semi} for rumor representation and classification, respectively.
PLAN~\cite{khoo2020plan} and RDLT~\cite{zhang2022rdlt} further extend these directed relationships to pairwise interactions using learnable attention mechanisms. Specifically, PLAN employs the attention mechanism in a transformer network to capture long-range relationships among comments, while RDLT uses it to enhance focus on long-tail comments. DDGCN~\cite{sun2022ddgcn} models the propagation dynamics with a graph convolution network that captures structural information at different time stages, and then combines the representations from different stages with a temporal fusion unit. CT-RvNN~\cite{peng2024ctrvnn} enhances the propagation model by incorporating comment time intervals as edge weights and constructs a coding tree~\cite{li2016structural} over the weighted propagation graph to extract essential structural information.

\subsubsection{Model Robustness}
In real-world scenarios, social media posts are often accompanied by unreliable responses, such as malicious comments designed to enhance the credibility of false information, and irrelevant noise comments unrelated to the source post. These factors significantly interfere with the accurate detection of rumors~\cite{yang2021rumor,ma2015detect,dataset_pheme9}. Therefore, enhancing the robustness of rumor detectors against noise, perturbations, and adversarial attacks, and developing models that can maintain high predictive performance in complex and noisy social media environments, is a crucial research direction.
EBGCN~\cite{wei2021ebgcn} models comment uncertainty by adaptively adjusting edge weights in propagation graphs from a probabilistic perspective, which are then utilized in the aggregation of GCNs. FGCN~\cite{wei2022fgcn} introduces a neuro-fuzzy method based on fuzzy theory~\cite{negoita1977fuzzy} to adapt edge weights according to predefined membership. DCE-RD~\cite{zhang23dcerd} designs a subgraph generation strategy to produce diverse counterfactual evidence, employing a Determinantal Point Process (DPP) based loss to enhance evidence diversity. The generated evidence is aggregated to provide robust veracity prediction. GARD~\cite{tao24gard} enhances detection robustness by capturing semantic evolution information through self-supervised learning, conducting feature reconstruction from both local and global perspectives. EIN~\cite{jiang2025ein} addresses sensitivity to data quality by integrating epidemiological knowledge~\cite{chang2023novel, wang2012simple}, modeling propagation dynamics and user stance with susceptible states, and utilizing LLMs to annotate comment stances. Similarly, JSDRV~\cite{yang2024reinforcement} employs LLMs for post-level stance annotation and develops a reinforcement learning-based model for joint stance detection and rumor veracity prediction.

Contrastive learning, a self-supervised approach, ensures graphs maintain similar embeddings under different perturbations by emphasizing invariant features, thereby reducing sensitivity to noise and enhancing model robustness and generalization~\cite{chen2020simple, he2020momentum}. This approach is also utilized to improve the robustness of rumor detectors. RDEA~\cite{he21rdea} pre-trains a GIN~\cite{xu2018powerful} encoder by maximizing mutual information between views under different perturbations, then fine-tunes the encoder for rumor veracity prediction. Node masking, edge dropping, and subgraph sampling operations are utilized for graph perturbation. RDCL~\cite{ma2022rdcl} further proposes six perturbation methods under node-based and topology-based categories, maximizing consistency between two perturbed graphs of the same original graph, and minimizing the distance between perturbed and original graphs from the same class. A hard positive strategy also enhances the performance of the contrastive learning framework. GACL~\cite{sun22gacl} utilizes contrastive learning to perceive differences between propagation graphs of the same and different classes, proposing an Adversarial Feature Transformation module to generate hard negative samples. CRFB~\cite{ma2023crfb} introduces a two-component (true-false) beta mixture model to distinguish true and false negative examples in contrastive learning, employing a CNN-based model to capture consistent and complementary information between two augmented propagation structures. RAGCL~\cite{cui2024ragcl} observes the wide and shallow characteristics of propagation trees and proposes augmenting original graphs following three principles: exempting root nodes, retaining deep reply nodes, and preserving lower-level nodes in deep sections, based on which node centrality-based importance scores are incorporated to generate augmented views. FADE~\cite{zhang2024fade} improves representation quality with contrastive learning among augmented graphs under an adaptive strategy, mitigating event bias by subtracting event-only predictions.

Recent studies have demonstrated the effectiveness of adversarial training in improving robustness against noise and attacks~\cite{zhao2024adversarial}. AARD~\cite{song21aard} considers both text content features and topology context when generating adversarial responses, utilizing the adversarial response generator to improve rumor detector robustness. GACL~\cite{sun22gacl} also employs adversarial training to produce conflicting samples and hard negative samples, enhancing the effectiveness and robustness of the rumor detector.

Generation-based methods are also employed to enhance rumor detector robustness under certain conditions. GenFEND~\cite{nan2024genfend} focuses on alleviating exposure bias and enhancing comment diversity, adopting large language models (LLMs) as user simulators and comment generators. KPG~\cite{dataset_weibo22} comprehensively considers both noisy propagation threads and spreading-limited rumors at early stages, proposing two interdependent modules to select key propagation graphs from generation-enhanced candidate graphs, utilizing a reinforcement learning framework to alternately update these modules. D$^2$~\cite{xu2025d2} aims to enhance early rumor detection by predicting possible diffusion paths based on limited propagation at early stages and users' social relationships.

Furthermore, some methods target improving transferability among rumors in different languages or topics. ACLR~\cite{dataset_weibo_covid19_twitter_covid19} overcomes domain and language restrictions via language alignment and a novel supervised contrastive training paradigm, developing an adversarial augmentation mechanism to enhance the robustness of low-resource rumor representation. To assess transferability under different test topics and languages, the Weibo-COVID19 dataset is utilized as the low-resource test set for evaluating proposed methods, while the Twitter15 and Twitter16 datasets serve as the well-resourced training set. Similarly, the Twitter-COVID19 dataset is used as the low-resource test set, and the Weibo dataset is employed as the well-resourced training set. T$^3$RD~\cite{zhang2024t3rd} introduces test-time self-supervised learning to enhance rumor detection performance on low-resource datasets, performing graph- and node-level contrastive learning as auxiliary tasks, and introducing feature alignment constraints to balance knowledge derived from the training set and test samples. FNDCD~\cite{gong2025fndcd} employs a reweighting strategy based on classification confidence and propagation structure regularization to reduce domain-specific biases, enhancing the detection of unseen fake news with new topics and domains.

\subsubsection{LLM-augmented Detection}
In recent years, there has been a growing interest in leveraging Large Language Models (LLMs) for rumor detection. Despite their remarkable capabilities in text understanding and generation, LLMs face significant challenges when dealing with complex structured information, particularly in effectively processing intricate topologies.
Specifically, LeRuD~\cite{liu2024can} designs specific prompts to guide the model's focus on crucial clues such as writing styles, commonsense errors, and rebuttals or conflicts in comments. It also divides the entire propagation information into a Chain-of-Propagation to alleviate the cognitive burden on LLMs. ARG~\cite{hu2024bad} prompts the LLM to generate multi-perspective rationales for rumors and utilizes smaller language models, such as BERT~\cite{devlin2019bert}, to integrate and infer rumor veracity from these rationales. SePro~\cite{zeng2025exploring} employs Graph Attention Networks~\cite{liu2018mining} and community detection methods~\cite{ng2001spectral} to identify and extract key subgraphs from the complex information network, thereby simplifying the structure that the LLM needs to process.

\section{Methods Based on Heterogeneous propagation}
\label{sec:hetero_prop}
\subsection{Problem Formulation}
\label{sec:prob2}
As introduced in Section~\ref{sec:prob1}, the task of rumor detection based on homogeneous propagation information involves predicting the veracity of the source post based on the commenting/retweeting interactions that form a propagation tree, making it a graph-level classification problem. In contrast, rumor detection with heterogeneous propagation incorporates a broader spectrum of information collected from social networks. This approach can be viewed as a node classification problem on heterogeneous graphs, where different types of nodes and edges represent various elements of the social context. Specifically, the social context typically includes: 1) users who published the source, comment, or retweet posts; 2) keywords extracted from the text content of the source news; and 3) semantically related posts discussing the same news event on other social platforms. To integrate all this information, a heterogeneous graph is constructed, with different types of nodes and edges representing the complex relationships within the social context.

Unlike rumor detection based on direct propagation information, which primarily focuses on the structure and dynamics of information spread, approaches with social context-based propagation introduce a higher level of complexity. Different methods in this category often adopt varied problem definitions, focusing on distinct aspects of the social context. To facilitate a comprehensive understanding, we abstract a foundational problem formulation that serves as a common ground for different variants of the problem.

Specifically, let ${S}$ denote the set of news posts, ${T}$ denote the set of comment/retweet posts interacting with the news posts in ${S}$, and ${U}$ denote the set of users on the social network. The set $\mathcal{U}$ includes users who directly interacted with the news posts (i.e., by publishing the news posts or replying to them with comments/retweets), as well as users indirectly associated with the news posts (i.e., having a friend relationship with direct users). These sets collectively form the different types of nodes in a heterogeneous graph, ${V}=S\cup T\cup U$. In this heterogeneous graph, the edges encompass various types of relationships, denoted by the edge set $E$. Specifically, edges between nodes in $S$ and $T$ indicate commenting or retweeting interactions. Edges between nodes in $U$ and $S$ represent users publishing news posts $s_i \in S$, or commenting on/retweeting news posts $s_i \in S$, while edges between $u_i \in U$ and $t_i\in T$ represent $u_i$ posting the comment/retweet post $t_i$. Additionally, edges between nodes in $U$ capture social relationships, such as interactions or friendships. Thus, the heterogeneous graph can be represented as $G=(V, E)$. The goal of this task is to learn a prediction function $f: \mathcal{S}\times \mathcal{G}\rightarrow \mathcal{Y}$, which classifies the news posts $s_i \in S$ into rumors (fake news) or non-rumors (real news) along with the heterogeneous social context propagation graph $G$, where $\mathcal{Y}$ is the space of veracity labels. The learning process is based on a set of labeled news posts $\mathcal{S}_{train}$, and the learned function $f$ is expected to accurately predict the veracity label of unlabeled news posts in $\mathcal{S}_{test}$: $\hat{y}_i=f(s_i, G)$. In some existing methods, only a subset of the node/edge types in the heterogeneous graph $G$ are utilized to enhance the effectiveness of rumor detection, while other approaches employ additional information to refine the types and relationships within $G$.

\subsection{Datasets}
\label{sec:dataset2}
In this section, we introduce the common datasets used by false information detection methods that leverage heterogeneous propagation. In addition to the Twitter15, Twitter16, and Weibo datasets introduced in Section~\ref{sec:dataset1}, the following datasets are also utilized:

\begin{itemize}
\item \textbf{Fakeddit}~\cite{nakamura2019r} is a large-scale benchmark dataset from Reddit\footnote{\url{https://www.reddit.com/}} that includes both text and image content. This dataset supports news classification at different granularities, including 2-class, 3-class, and 6-class classifications. Specifically, the 2-way classification determines whether a sample is fake or true. The 3-way classification distinguishes between completely true samples, direct quotes from propaganda posters, and fake samples. The 6-way classification includes the following labels: \textit{True, Satire/Parody, Misleading Content, Imposter Content, False Connection}, and \textit{Manipulated Content}.

\item \textbf{GossipCop}~\cite{shu2020fakenewsnet} collects news from GossipCop\footnote{\url{https://www.gossipcop.com/}}, a website that provides rating scores for entertainment stories on a scale from 0 to 10, reflecting the degree from fake to real. The dataset includes stories with scores under 5 as fake news, and collects related discussion posts on Twitter, along with their commenting/retweeting interactions, engaged user profiles, user friendship information, and spatial locations provided in user profiles.

\item \textbf{PolitiFact}~\cite{shu2020fakenewsnet} gathers news content based on annotations from PolitiFact\footnote{\url{https://www.politifact.com/}}, a fact-checking website where political news is evaluated as fake or real by journalists and domain experts. Similar to the GossipCop dataset, the PolitiFact dataset also includes semantically related tweet posts on Twitter, along with rich propagation information on social contexts, including both comment/retweet interactions and comprehensive user information associated with the posts.

\item \textbf{FakeHealth}~\cite{dai2020ginger} is collected from the healthcare information review website, Health News Review\footnote{\url{https://www.healthnewsreview.org/}}, and contains over 2000 news articles, 500k posts, and 27k user profiles, along with user networks.

\item \textbf{Mc-Fake}~\cite{min2022divide} comprises labeled fake news from existing datasets and real news from reliable sources, covering five topics: \textit{Politics, Entertainment, Health, Covid-19}, and \textit{Syria War}. Related tweets, retweets, and replies, along with the corresponding users on the Twitter platform, are retrieved, forming a dataset with rich social contexts.

\end{itemize}

\subsection{Method Review}
In this section, we compile and review existing methods for false information detection that leverage heterogeneous propagation information on social networks. Unlike the methods discussed in Section~\ref{sec:homo_prop}, which focus on homogeneous propagation patterns, the methods reviewed here incorporate a broader range of contextual information available on social networks. Specifically, we classify these methods into two main categories based on the types of additional context they utilize: \textit{1)} user-related context and \textit{2)} other context.

\subsubsection{User-Related Context}
In this category, methods collect information about engaged users as social context, providing complementary evidence to assist in detecting false information. The user context not only provides connections across different news events, implying implicit information dissemination, but also reflects user preferences through user profiles and friendship relationships, offering useful clues for veracity classification.
Specifically, GLAN~\cite{yuan2019jointly} integrates user information to model local and global propagation on social networks using a heterogeneous graph, involving news posts, commenting posts, and the users publishing these posts. An attention mechanism is employed to enhance the representation of news posts by fusing the semantic information in the constructed graph.
SMAN~\cite{yuan2020early} further categorizes users into two types: 1) publishers, the authors of news posts, and 2) comment users who post comments on news posts. Additionally, SMAN annotates the credibility of a subset of users based on their historical publishing and reposting behaviors as supervised information, and jointly predicts rumor veracity and user credibility.
FANG~\cite{nguyen2020fang} follows this user type refinement and proposes an inductive graph learning framework~\cite{hamilton2017inductive} that produces representations effectively and efficiently.
UPFD~\cite{dou2021user} is inspired by the confirmation bias theory, which suggests that the probability of users spreading fake news is related to their preferences, which can be extracted from users' historical posts. Specifically, UPFD models the target news and user historical post contents, extracts user-user interaction information, and fuses these endogenous preferences and exogenous context together.
DUCK~\cite{tian2022duck} extracts information from the user-user interaction network, the post-post commenting network, and conversation threads using Graph Attention Networks (GAT)~\cite{vieweg2010microblogging}, BERT~\cite{devlin2019bert} combined with GAT, and transformers~\cite{beltagy2020longformer}.
PSIN~\cite{min2022divide} applies a divide-and-conquer strategy to model the heterogeneous relations and employs three variants of Graph Attention Networks to model post-post, user-user, and post-user subgraphs. An additional adversarial topic discriminator is adopted to learn topic-agnostic features for enhanced veracity prediction.
DECOR~\cite{wu2023decor} computes the number of common engaged users between news articles to form a news engagement graph. DECOR further reveals that the degree distributions of fake and real news in this engagement graph show a clear difference, and proposes a Degree-Corrected Stochastic Block model to enhance fake news detection.
PSGT~\cite{zhu2024propagation} focuses on user-user interaction relationships, proposing a noise-reduction self-attention mechanism based on the information bottleneck principle and a novel relational propagation graph as a position encoding for the graph Transformer.

Special learning settings, such as adversarial attack and transfer learning, are also explored. For instance, QSA-AC~\cite{wang2024bots} considers attacking the user-news engagement network by injecting bots into the network to obfuscate GNN-based fake news detectors. QSA-AC follows a surrogate-based approach, leveraging the query process to enhance the effectiveness of the attack. Meanwhile, MMHT~\cite{yang2025macro} utilizes user engagement and historical preference information for cross-domain fake news detection. Specifically, MMHT disentangles news content and user engagement from a macro perspective and disentangles veracity-relevant and veracity-irrelevant features from a micro perspective. These features are then utilized to facilitate more effective knowledge transfer.

\subsubsection{Other Context}
In addition to leveraging user information as social context, methods in this category broaden the scope of social context by incorporating various types of information. For instance, HDGCN~\cite{kang2021fake} collects the temporal and domain information of news articles and proposes a neighbor sampling strategy along with a novel hierarchical attention mechanism to enhance news embedding and classification. DiHAN~\cite{chang2024dihan} constructs a heterogeneous graph that includes the core subjects of the news and their temporal information. It extracts meta-paths considering both node/edge types and the chronological interactions between news items, employing a hierarchical attention mechanism to integrate different types of meta-path-based temporal information.
Temporal information is also utilized in DWAN~\cite{kim2025revisiting} for train-test splitting, introducing a new approach where the model is trained on earlier news and tested on subsequent news. DWAN improves detection performance under this temporality-aware setting by using a Graph Structure Learning framework to reassign weights to existing edges based on a learned edge weight estimator.
FinerFact~\cite{jin2022towards} mines social context information through semantic relations to enable fine-grained reasoning. Specifically, FinerFact extracts keywords from news articles and posts on other social media platforms, using them as connectors between news, discussion posts, and the users who publish these posts. This forms a heterogeneous graph rich in social contextual information. The credibility of users is combined with the graph topology to reason about the veracity of rumors.
Following this graph construction approach, SureFact~\cite{yang2022reinforcement} proposes a reinforced subgraph generation method and develops a Hierarchical Path-aware Kernel Graph Attention Network to reason about rumor veracity based on the generated subgraphs. This subgraph reasoning paradigm not only provides clear interpretability of the prediction results but also enhances generalization.

\section{Future Work}
\label{sec:future}
\subsection{Development of a Comprehensive Benchmark for False Information Detection}
The field of false information detection currently lacks a practical, off-the-shelf benchmark that can standardize research and facilitate direct comparisons. This deficiency is particularly evident in three critical aspects: unified train-validation-test splitting, standardized evaluation process, and unified data platform. 

Firstly, in existing studies, the train-test split is often conducted independently by different researchers, leading to non-comparable outcomes across studies and inconsistency with real-world scenarios~\cite{kim2025revisiting}. To address these challenges, we propose the development of a benchmark that includes commonly used datasets and provides a unified train-validation-test split. This split should be based on meaningful dimensions, such as topic or time, to better simulate real-world conditions and assess the robustness and generalizability of detection methods.
For instance, splitting the dataset by topic or time stamp, with the training and test sets containing entirely different topics, can effectively evaluate a method's ability to generalize across diverse subjects. This approach also allows researchers to analyze how generalization performance varies with different topics.

Secondly, another critical gap in current research is the lack of a unified evaluation process~\cite{dataset_weibo22}. Different methods often employ distinct evaluation codes and adopt various metrics, making it difficult to compare their performance directly. A standardized benchmark should include a consistent evaluation protocol, with clearly defined metrics and evaluation scripts. This will ensure that all methods are assessed under the same conditions, facilitating fair comparisons and promoting the development of more effective detection techniques.

Thirdly, the dispersion of datasets across different studies also poses challenges for researchers. A comprehensive, off-the-shelf benchmark platform that integrates multiple datasets would significantly facilitate research in this area. Such a platform would provide a one-stop solution for researchers to access and work with standardized datasets, reducing the time and effort required to prepare data and allowing them to focus more on developing and testing new methods. This centralized approach would also promote collaboration and reproducibility, driving the field forward more efficiently.

\subsection{Diversifying Information Carriers for False Information Detection}
Current propagation-based false information detection primarily focuses on news or microblogs, where text is the dominant information carrier, occasionally supplemented by images. 
However, the landscape of misinformation is evolving beyond text-centric media. The rapid growth of platforms for short videos, podcasts, and other multimedia content has made them prevalent and concerning channels for disseminating false information. 
Taking short videos as a prominent example, their inherent multimodal complexity and unique propagation dynamics pose a significant challenge to existing detection techniques. This calls for dedicated research into novel methodologies that can jointly analyze content and propagation patterns, as well as the creation of comprehensive datasets to address this evolving problem~\cite{wu2024interpretable, yang2023multimodal}.

\subsection{Mitigating the Impact of False Information}
Detecting and removing false information from social platforms (e.g., by deleting posts) is a straightforward approach to mitigate the impact of false information. However, this method has several drawbacks. First, it may mistakenly remove content that is not false information, thereby affecting user experience. Second, it does not address the root causes of why people believe and propagate false information. Finally, this method heavily relies on platform operations and may not comprehensively solve the problem of false information. Therefore, how to effectively reduce the impact of potentially false information on social networks is a critical issue that requires in-depth research.

One potential solution is to utilize bots~\cite{hofeditz2019meaningful,hayawi2023social,ng2025global} on social platforms to disseminate corrective information. This approach can automatically push accurate information to users once false information is detected, thereby mitigating the spread of misinformation. However, several key aspects need to be considered for this method to be effective. The content of corrective information needs to be based on both the original information and the propagated information. This is because the propagated information reflects the community's understanding of the original message, and generating corrective posts from this understanding can increase the likelihood of acceptance. Furthermore, the dissemination strategy for corrective information should be precisely tailored by integrating the propagation information of the original post and relevant social context information.

Despite its potential, this approach currently faces many challenges. The problem definition and evaluation of such a correction mechanism are still unclear. For example, how to measure the effectiveness of corrective information and determine its role in reducing the spread of false information remains an open question. Moreover, there is a lack of high-quality datasets for researching this issue. Existing datasets mostly focus on the detection of false information, while data related to the generation, dissemination, and user feedback of corrective information are relatively scarce. This makes it difficult for researchers to systematically study and validate relevant methods.

\section{Conclusion}
This study systematically reviews existing false information detection methods through the lens of information propagation on social networks. We begin by categorizing information dissemination into homogeneous and heterogeneous propagation. Based on this categorization, we introduce a comprehensive taxonomy that encompasses existing detection methods. We then provide a detailed overview and summary of related methods within these two categories, covering formal problem formulations, popular datasets, and detailed approaches, with an emphasis on the utilization of propagation dynamics and social context. Furthermore, we identify several promising avenues for future research, including the development of standardized evaluation frameworks, innovative strategies to mitigate the impact of false information, and the exploration of diverse information carriers.

\section*{Acknowledgements}
This work is supported by the RGC GRF grant (No. 14217322) and the Tencent WeChat Rhino-Bird Focused Research Program.

\renewcommand{\bibsep}{1pt}
\renewcommand{\bibfont}{\small}
\bibliographystyle{plainnat}
\bibliography{ref}

\begin{thebibliography}{98}
\providecommand{\natexlab}[1]{#1}
\providecommand{\url}[1]{\texttt{#1}}
\expandafter\ifx\csname urlstyle\endcsname\relax
  \providecommand{\doi}[1]{doi: #1}\else
  \providecommand{\doi}{doi: \begingroup \urlstyle{rm}\Url}\fi

\bibitem[Bakdash et~al.(2018)Bakdash, Sample, Rankin, Kantarcioglu, Holmes, Kase, Zaroukian, and Szymanski]{bakdash2018future}
Jonathan Bakdash, Char Sample, Monica Rankin, Murat Kantarcioglu, Jennifer Holmes, Sue Kase, Erin Zaroukian, and Boleslaw Szymanski.
\newblock The future of deception: Machine-generated and manipulated images, video, and audio?
\newblock In \emph{2018 International Workshop on Social Sensing (SocialSens)}, pages 2--2. IEEE, 2018.

\bibitem[Battista et~al.(2025)Battista, Lanciano, and Curci]{battista2025survey}
Fabiana Battista, Tiziana Lanciano, and Antonietta Curci.
\newblock A survey on the criteria used to judge (fake) news in italian population.
\newblock \emph{Brain and Behavior}, 15\penalty0 (2):\penalty0 e70315, 2025.

\bibitem[Beltagy et~al.(2020)Beltagy, Peters, and Cohan]{beltagy2020longformer}
Iz~Beltagy, Matthew~E Peters, and Arman Cohan.
\newblock Longformer: The long-document transformer.
\newblock \emph{arXiv preprint arXiv:2004.05150}, 2020.

\bibitem[Bian et~al.(2020)Bian, Xiao, Xu, Zhao, Huang, Rong, and Huang]{bian2020bigcn}
Tian Bian, Xi~Xiao, Tingyang Xu, Peilin Zhao, Wenbing Huang, Yu~Rong, and Junzhou Huang.
\newblock Rumor detection on social media with bi-directional graph convolutional networks.
\newblock In \emph{{AAAI}}, pages 549--556, 2020.

\bibitem[Chang et~al.(2024)Chang, Hu, Li, Yang, Jiang, and Sun]{chang2024dihan}
Ya-Ting Chang, Zhibo Hu, Xiaoyu Li, Shuiqiao Yang, Jiaojiao Jiang, and Nan Sun.
\newblock Dihan: A novel dynamic hierarchical graph attention network for fake news detection.
\newblock In \emph{Proceedings of the 33rd ACM International Conference on Information and Knowledge Management}, pages 197--206, 2024.

\bibitem[Chang and de~Jong(2023)]{chang2023novel}
You Chang and Mart~CM de~Jong.
\newblock A novel method to jointly estimate transmission rate and decay rate parameters in environmental transmission models.
\newblock \emph{Epidemics}, 42:\penalty0 100672, 2023.

\bibitem[Chen et~al.(2020)Chen, Kornblith, Norouzi, and Hinton]{chen2020simple}
Ting Chen, Simon Kornblith, Mohammad Norouzi, and Geoffrey Hinton.
\newblock A simple framework for contrastive learning of visual representations.
\newblock In \emph{International conference on machine learning}, pages 1597--1607. PmLR, 2020.

\bibitem[Cho et~al.(2014)Cho, Van~Merri{\"e}nboer, Bahdanau, and Bengio]{cho2014properties}
Kyunghyun Cho, Bart Van~Merri{\"e}nboer, Dzmitry Bahdanau, and Yoshua Bengio.
\newblock On the properties of neural machine translation: Encoder-decoder approaches.
\newblock \emph{arXiv preprint arXiv:1409.1259}, 2014.

\bibitem[Cui and Jia(2024)]{cui2024ragcl}
Chaoqun Cui and Caiyan Jia.
\newblock Propagation tree is not deep: Adaptive graph contrastive learning approach for rumor detection.
\newblock In \emph{AAAI}, pages 73--81, 2024.

\bibitem[Dai et~al.(2020)Dai, Sun, and Wang]{dai2020ginger}
Enyan Dai, Yiwei Sun, and Suhang Wang.
\newblock Ginger cannot cure cancer: Battling fake health news with a comprehensive data repository.
\newblock In \emph{Proceedings of the International AAAI Conference on Web and Social Media}, volume~14, pages 853--862, 2020.

\bibitem[Devlin et~al.(2019)Devlin, Chang, Lee, and Toutanova]{devlin2019bert}
Jacob Devlin, Ming-Wei Chang, Kenton Lee, and Kristina Toutanova.
\newblock Bert: Pre-training of deep bidirectional transformers for language understanding.
\newblock In \emph{Proceedings of the 2019 conference of the North American chapter of the association for computational linguistics: human language technologies, volume 1 (long and short papers)}, pages 4171--4186, 2019.

\bibitem[Dou et~al.(2021)Dou, Shu, Xia, Yu, and Sun]{dou2021user}
Yingtong Dou, Kai Shu, Congying Xia, Philip~S Yu, and Lichao Sun.
\newblock User preference-aware fake news detection.
\newblock In \emph{Proceedings of the 44th international ACM SIGIR conference on research and development in information retrieval}, pages 2051--2055, 2021.

\bibitem[Fallis(2014)]{fallis2014functional}
Don Fallis.
\newblock A functional analysis of disinformation.
\newblock \emph{IConference 2014 Proceedings}, 2014.

\bibitem[Gong et~al.(2023)Gong, Sinnott, Qi, and Paris]{gong2023fake}
Shuzhi Gong, Richard~O Sinnott, Jianzhong Qi, and Cecile Paris.
\newblock Fake news detection through graph-based neural networks: A survey.
\newblock \emph{arXiv preprint arXiv:2307.12639}, 2023.

\bibitem[Gong et~al.(2025)Gong, Sinnott, Qi, and Paris]{gong2025fndcd}
Shuzhi Gong, Richard Sinnott, Jianzhong Qi, and Cecile Paris.
\newblock Unseen fake news detection through casual debiasing.
\newblock In \emph{Companion Proceedings of the ACM on Web Conference 2025}, pages 981--985, 2025.

\bibitem[Guo et~al.(2020)Guo, Ding, Yao, Liang, and Yu]{guo2020future}
Bin Guo, Yasan Ding, Lina Yao, Yunji Liang, and Zhiwen Yu.
\newblock The future of false information detection on social media: New perspectives and trends.
\newblock \emph{ACM Computing Surveys (CSUR)}, 53\penalty0 (4):\penalty0 1--36, 2020.

\bibitem[Hamilton et~al.(2017)Hamilton, Ying, and Leskovec]{hamilton2017inductive}
Will Hamilton, Zhitao Ying, and Jure Leskovec.
\newblock Inductive representation learning on large graphs.
\newblock \emph{Advances in neural information processing systems}, 30, 2017.

\bibitem[Hayawi et~al.(2023)Hayawi, Saha, Masud, Mathew, and Kaosar]{hayawi2023social}
Kadhim Hayawi, Susmita Saha, Mohammad~Mehedy Masud, Sujith~Samuel Mathew, and Mohammed Kaosar.
\newblock Social media bot detection with deep learning methods: a systematic review.
\newblock \emph{Neural Computing and Applications}, 35\penalty0 (12):\penalty0 8903--8918, 2023.

\bibitem[He et~al.(2020)He, Fan, Wu, Xie, and Girshick]{he2020momentum}
Kaiming He, Haoqi Fan, Yuxin Wu, Saining Xie, and Ross Girshick.
\newblock Momentum contrast for unsupervised visual representation learning.
\newblock In \emph{Proceedings of the IEEE/CVF conference on computer vision and pattern recognition}, pages 9729--9738, 2020.

\bibitem[He et~al.(2021)He, Li, Zhou, and Yang]{he21rdea}
Zhenyu He, Ce~Li, Fan Zhou, and Yi~Yang.
\newblock Rumor detection on social media with event augmentations.
\newblock In \emph{{SIGIR}}, pages 2020--2024, 2021.

\bibitem[Hofeditz et~al.(2019)Hofeditz, Ehnis, Bunker, Brachten, and Stieglitz]{hofeditz2019meaningful}
Lennart Hofeditz, Christian Ehnis, Deborah Bunker, Florian Brachten, and Stefan Stieglitz.
\newblock Meaningful use of social bots? possible applications in crisis communication during disasters.
\newblock In \emph{ECIS}, pages 1--16, 2019.

\bibitem[Hu et~al.(2024)Hu, Sheng, Cao, Shi, Li, Wang, and Qi]{hu2024bad}
Beizhe Hu, Qiang Sheng, Juan Cao, Yuhui Shi, Yang Li, Danding Wang, and Peng Qi.
\newblock Bad actor, good advisor: Exploring the role of large language models in fake news detection.
\newblock In \emph{Proceedings of the AAAI Conference on Artificial Intelligence}, volume~38, pages 22105--22113, 2024.

\bibitem[Hussain et~al.(2025)Hussain, Wasim, Hameed, Rehman, Asim, and Dengel]{hussain2025fake}
Fiza~Gulzar Hussain, Muhammad Wasim, Seemab Hameed, Abdur Rehman, Muhammad~Nabeel Asim, and Andreas Dengel.
\newblock Fake news detection landscape: Datasets, data modalities, ai approaches, their challenges, and future perspectives.
\newblock \emph{IEEE Access}, 2025.

\bibitem[Jiang et~al.(2025)Jiang, Chen, Gao, Zhang, Cui, and Yin]{jiang2025ein}
Wei Jiang, Tong Chen, Xinyi Gao, Wentao Zhang, Lizhen Cui, and Hongzhi Yin.
\newblock Epidemiology-informed network for robust rumor detection.
\newblock In \emph{Proceedings of the ACM on Web Conference 2025}, pages 3618--3627, 2025.

\bibitem[Jin et~al.(2022)Jin, Wang, Yang, Sun, Wang, Liao, and Xie]{jin2022towards}
Yiqiao Jin, Xiting Wang, Ruichao Yang, Yizhou Sun, Wei Wang, Hao Liao, and Xing Xie.
\newblock Towards fine-grained reasoning for fake news detection.
\newblock In \emph{Proceedings of the AAAI Conference on Artificial Intelligence}, volume~36, pages 5746--5754, 2022.

\bibitem[Jin et~al.(2017)Jin, Cao, Guo, Zhang, Wang, and Luo]{jin2017detection}
Zhiwei Jin, Juan Cao, Han Guo, Yongdong Zhang, Yu~Wang, and Jiebo Luo.
\newblock Detection and analysis of 2016 us presidential election related rumors on twitter.
\newblock In \emph{Social, Cultural, and Behavioral Modeling: 10th International Conference, SBP-BRiMS 2017, Washington, DC, USA, July 5-8, 2017, Proceedings 10}, pages 14--24. Springer, 2017.

\bibitem[Kang et~al.(2021)Kang, Cao, Shang, Liang, Tang, and Tong]{kang2021fake}
Zhezhou Kang, Yanan Cao, Yanmin Shang, Tao Liang, Hengzhu Tang, and Lingling Tong.
\newblock Fake news detection with heterogenous deep graph convolutional network.
\newblock In \emph{Pacific-Asia Conference on Knowledge Discovery and Data Mining}, pages 408--420. Springer, 2021.

\bibitem[Kar et~al.(2021)Kar, Bhardwaj, Samanta, and Azad]{kar2021no}
Debanjana Kar, Mohit Bhardwaj, Suranjana Samanta, and Amar~Prakash Azad.
\newblock No rumours please! a multi-indic-lingual approach for covid fake-tweet detection.
\newblock In \emph{2021 grace hopper celebration India (GHCI)}, pages 1--5. IEEE, 2021.

\bibitem[Khoo et~al.(2020)Khoo, Chieu, Qian, and Jiang]{khoo2020plan}
Ling Min~Serena Khoo, Hai~Leong Chieu, Zhong Qian, and Jing Jiang.
\newblock Interpretable rumor detection in microblogs by attending to user interactions.
\newblock In \emph{Proceedings of the AAAI conference on artificial intelligence}, volume~34, pages 8783--8790, 2020.

\bibitem[Kim et~al.(2025)Kim, Lee, In, Yoon, and Park]{kim2025revisiting}
Junghoon Kim, Junmo Lee, Yeonjun In, Kanghoon Yoon, and Chanyoung Park.
\newblock Revisiting fake news detection: Towards temporality-aware evaluation by leveraging engagement earliness.
\newblock In \emph{Proceedings of the Eighteenth ACM International Conference on Web Search and Data Mining}, pages 559--567, 2025.

\bibitem[Kipf and Welling(2016)]{kipf2016semi}
Thomas~N Kipf and Max Welling.
\newblock Semi-supervised classification with graph convolutional networks.
\newblock \emph{arXiv preprint arXiv:1609.02907}, 2016.

\bibitem[Kumar and Shah(2018)]{kumar2018false}
Srijan Kumar and Neil Shah.
\newblock False information on web and social media: A survey.
\newblock \emph{arXiv preprint arXiv:1804.08559}, 2018.

\bibitem[Kwao et~al.(2025)Kwao, Yang, Zou, and Ma]{kwao2025survey}
Lazarus Kwao, Yang Yang, Jie Zou, and Jing Ma.
\newblock A survey of approaches to early rumor detection on microblogging platforms: Computational and socio-psychological insights.
\newblock \emph{Wiley Interdisciplinary Reviews: Data Mining and Knowledge Discovery}, 15\penalty0 (1):\penalty0 e70001, 2025.

\bibitem[Lakzaei et~al.(2024)Lakzaei, Haghir~Chehreghani, and Bagheri]{lakzaei2024disinformation}
Batool Lakzaei, Mostafa Haghir~Chehreghani, and Alireza Bagheri.
\newblock Disinformation detection using graph neural networks: a survey.
\newblock \emph{Artificial Intelligence Review}, 57\penalty0 (3):\penalty0 52, 2024.

\bibitem[Lazer et~al.(2018)Lazer, Baum, Benkler, Berinsky, Greenhill, Menczer, Metzger, Nyhan, Pennycook, Rothschild, et~al.]{lazer2018science}
David~MJ Lazer, Matthew~A Baum, Yochai Benkler, Adam~J Berinsky, Kelly~M Greenhill, Filippo Menczer, Miriam~J Metzger, Brendan Nyhan, Gordon Pennycook, David Rothschild, et~al.
\newblock The science of fake news.
\newblock \emph{Science}, 359\penalty0 (6380):\penalty0 1094--1096, 2018.

\bibitem[Li and Pan(2016)]{li2016structural}
Angsheng Li and Yicheng Pan.
\newblock Structural information and dynamical complexity of networks.
\newblock \emph{IEEE Transactions on Information Theory}, 62\penalty0 (6):\penalty0 3290--3339, 2016.

\bibitem[Li et~al.(2025)Li, Qiao, Yin, Wu, Gao, Wang, and Li]{li2025survey}
Xianghua Li, Jiao Qiao, Shu Yin, Lianwei Wu, Chao Gao, Zhen Wang, and Xuelong Li.
\newblock A survey of multimodal fake news detection: A cross-modal interaction perspective.
\newblock \emph{IEEE Transactions on Emerging Topics in Computational Intelligence}, 2025.

\bibitem[Lin et~al.(2022)Lin, Ma, Chen, Yang, Cheng, and Chen]{dataset_weibo_covid19_twitter_covid19}
Hongzhan Lin, Jing Ma, Liangliang Chen, Zhiwei Yang, Mingfei Cheng, and Guang Chen.
\newblock Detect rumors in microblog posts for low-resource domains via adversarial contrastive learning.
\newblock \emph{arXiv preprint arXiv:2204.08143}, 2022.

\bibitem[Liu et~al.(2018)Liu, Yu, Wu, and Wang]{liu2018mining}
Qiang Liu, Feng Yu, Shu Wu, and Liang Wang.
\newblock Mining significant microblogs for misinformation identification: an attention-based approach.
\newblock \emph{ACM Transactions on Intelligent Systems and Technology (TIST)}, 9\penalty0 (5):\penalty0 1--20, 2018.

\bibitem[Liu et~al.(2024)Liu, Tao, Wu, Wu, and Wang]{liu2024can}
Qiang Liu, Xiang Tao, Junfei Wu, Shu Wu, and Liang Wang.
\newblock Can large language models detect rumors on social media?
\newblock \emph{arXiv preprint arXiv:2402.03916}, 2024.

\bibitem[Liu et~al.(2015)Liu, Nourbakhsh, Li, Fang, and Shah]{liu2015real}
Xiaomo Liu, Armineh Nourbakhsh, Quanzhi Li, Rui Fang, and Sameena Shah.
\newblock Real-time rumor debunking on twitter.
\newblock In \emph{Proceedings of the 24th ACM international on conference on information and knowledge management}, pages 1867--1870, 2015.

\bibitem[Ma et~al.(2022)Ma, Hu, Ge, Chen, Zhang, and Zhang]{ma2022rdcl}
Guanghui Ma, Chunming Hu, Ling Ge, Junfan Chen, Hong Zhang, and Richong Zhang.
\newblock Towards robust false information detection on social networks with contrastive learning.
\newblock In \emph{Proceedings of the 31st ACM international conference on information \& knowledge management}, pages 1441--1450, 2022.

\bibitem[Ma et~al.(2023)Ma, Dai, Liu, Han, and Ai]{ma2023crfb}
Jiachen Ma, Jing Dai, Yong Liu, Meng Han, and Chunyu Ai.
\newblock Contrastive learning for rumor detection via fitting beta mixture model.
\newblock In \emph{Proceedings of the 32nd ACM International Conference on Information and Knowledge Management}, pages 4160--4164, 2023.

\bibitem[Ma et~al.(2015)Ma, Gao, Wei, Lu, and Wong]{ma2015detect}
Jing Ma, Wei Gao, Zhongyu Wei, Yueming Lu, and Kam-Fai Wong.
\newblock Detect rumors using time series of social context information on microblogging websites.
\newblock In \emph{Proceedings of the 24th ACM international on conference on information and knowledge management}, pages 1751--1754, 2015.

\bibitem[Ma et~al.(2016)Ma, Gao, Mitra, Kwon, Jansen, Wong, and Cha]{dataset_weibo}
Jing Ma, Wei Gao, Prasenjit Mitra, Sejeong Kwon, Bernard~J Jansen, Kam-Fai Wong, and Meeyoung Cha.
\newblock Detecting rumors from microblogs with recurrent neural networks.
\newblock 2016.

\bibitem[Ma et~al.(2017)Ma, Gao, and Wong]{dataset_twitter15_twitter16}
Jing Ma, Wei Gao, and Kam-Fai Wong.
\newblock Detect rumors in microblog posts using propagation structure via kernel learning.
\newblock Association for Computational Linguistics, 2017.

\bibitem[Ma et~al.(2018)Ma, Gao, and Wong]{ma2018rvnn}
Jing Ma, Wei Gao, and Kam-Fai Wong.
\newblock Rumor detection on twitter with tree-structured recursive neural networks.
\newblock Association for Computational Linguistics, 2018.

\bibitem[Mahdi and Shati(2024)]{mahdi2024survey}
Alaa~Safaa Mahdi and Narjis~Mezaal Shati.
\newblock A survey on fake news detection in social media using graph neural networks.
\newblock \emph{Journal of Al-Qadisiyah for Computer Science and Mathematics}, 16\penalty0 (2):\penalty0 23--41, 2024.

\bibitem[Min et~al.(2022)Min, Rong, Bian, Xu, Zhao, Huang, and Ananiadou]{min2022divide}
Erxue Min, Yu~Rong, Yatao Bian, Tingyang Xu, Peilin Zhao, Junzhou Huang, and Sophia Ananiadou.
\newblock Divide-and-conquer: Post-user interaction network for fake news detection on social media.
\newblock In \emph{Proceedings of the ACM web conference 2022}, pages 1148--1158, 2022.

\bibitem[Mostafa et~al.(2024)Mostafa, Almogren, Al-Qurishi, and Alrubaian]{mostafa2024modality}
Mohamed Mostafa, Ahmad~S Almogren, Muhammad Al-Qurishi, and Majed Alrubaian.
\newblock Modality deep-learning frameworks for fake news detection on social networks: a systematic literature review.
\newblock \emph{ACM Computing Surveys}, 57\penalty0 (3):\penalty0 1--50, 2024.

\bibitem[Nakamura et~al.(2019)Nakamura, Levy, and Wang]{nakamura2019r}
Kai Nakamura, Sharon Levy, and William~Yang Wang.
\newblock r/fakeddit: A new multimodal benchmark dataset for fine-grained fake news detection.
\newblock \emph{arXiv preprint arXiv:1911.03854}, 2019.

\bibitem[Nan et~al.(2021)Nan, Cao, Zhu, Wang, and Li]{dataset_weibo21}
Qiong Nan, Juan Cao, Yongchun Zhu, Yanyan Wang, and Jintao Li.
\newblock Mdfend: Multi-domain fake news detection.
\newblock In \emph{Proceedings of the 30th ACM international conference on information \& knowledge management}, pages 3343--3347, 2021.

\bibitem[Nan et~al.(2024)Nan, Sheng, Cao, Hu, Wang, and Li]{nan2024genfend}
Qiong Nan, Qiang Sheng, Juan Cao, Beizhe Hu, Danding Wang, and Jintao Li.
\newblock Let silence speak: Enhancing fake news detection with generated comments from large language models.
\newblock In \emph{Proceedings of the 33rd ACM International Conference on Information and Knowledge Management}, pages 1732--1742, 2024.

\bibitem[Negoita and Ralescu(1977)]{negoita1977fuzzy}
CV~Negoita and DA~Ralescu.
\newblock On fuzzy optimization.
\newblock \emph{Kybernetes}, 6\penalty0 (3):\penalty0 193--195, 1977.

\bibitem[Ng et~al.(2001)Ng, Jordan, and Weiss]{ng2001spectral}
Andrew Ng, Michael Jordan, and Yair Weiss.
\newblock On spectral clustering: Analysis and an algorithm.
\newblock \emph{Advances in neural information processing systems}, 14, 2001.

\bibitem[Ng and Carley(2025)]{ng2025global}
Lynnette Hui~Xian Ng and Kathleen~M Carley.
\newblock A global comparison of social media bot and human characteristics.
\newblock \emph{Scientific Reports}, 15\penalty0 (1):\penalty0 10973, 2025.

\bibitem[Nguyen et~al.(2020)Nguyen, Sugiyama, Nakov, and Kan]{nguyen2020fang}
Van-Hoang Nguyen, Kazunari Sugiyama, Preslav Nakov, and Min-Yen Kan.
\newblock Fang: Leveraging social context for fake news detection using graph representation.
\newblock In \emph{Proceedings of the 29th ACM international conference on information \& knowledge management}, pages 1165--1174, 2020.

\bibitem[Peng et~al.(2024)Peng, Wu, Liu, and Xu]{peng2024ctrvnn}
Xingyu Peng, Junran Wu, Ruomei Liu, and Ke~Xu.
\newblock Rumor detection on social media with temporal propagation structure optimization.
\newblock \emph{arXiv preprint arXiv:2412.08316}, 2024.

\bibitem[Phan et~al.(2023)Phan, Nguyen, and Hwang]{phan2023fake}
Huyen~Trang Phan, Ngoc~Thanh Nguyen, and Dosam Hwang.
\newblock Fake news detection: A survey of graph neural network methods.
\newblock \emph{Applied Soft Computing}, 139:\penalty0 110235, 2023.

\bibitem[Shu et~al.(2017)Shu, Sliva, Wang, Tang, and Liu]{shu2017fake}
Kai Shu, Amy Sliva, Suhang Wang, Jiliang Tang, and Huan Liu.
\newblock Fake news detection on social media: A data mining perspective.
\newblock \emph{ACM SIGKDD explorations newsletter}, 19\penalty0 (1):\penalty0 22--36, 2017.

\bibitem[Shu et~al.(2020)Shu, Mahudeswaran, Wang, Lee, and Liu]{shu2020fakenewsnet}
Kai Shu, Deepak Mahudeswaran, Suhang Wang, Dongwon Lee, and Huan Liu.
\newblock Fakenewsnet: A data repository with news content, social context, and spatiotemporal information for studying fake news on social media.
\newblock \emph{Big data}, 8\penalty0 (3):\penalty0 171--188, 2020.

\bibitem[Song et~al.(2021)Song, Chen, Chang, Weng, and Shuai]{song21aard}
Yun{-}Zhu Song, Yi{-}Syuan Chen, Yi{-}Ting Chang, Shao{-}Yu Weng, and Hong{-}Han Shuai.
\newblock Adversary-aware rumor detection.
\newblock In \emph{{ACL/IJCNLP}}, pages 1371--1382, 2021.

\bibitem[Sun et~al.(2022{\natexlab{a}})Sun, Zhang, Zheng, and Ma]{sun2022ddgcn}
Mengzhu Sun, Xi~Zhang, Jiaqi Zheng, and Guixiang Ma.
\newblock Ddgcn: Dual dynamic graph convolutional networks for rumor detection on social media.
\newblock In \emph{Proceedings of the AAAI conference on artificial intelligence}, volume~36, pages 4611--4619, 2022{\natexlab{a}}.

\bibitem[Sun et~al.(2022{\natexlab{b}})Sun, Qian, Dong, Li, and Zhu]{sun22gacl}
Tiening Sun, Zhong Qian, Sujun Dong, Peifeng Li, and Qiaoming Zhu.
\newblock Rumor detection on social media with graph adversarial contrastive learning.
\newblock In \emph{{WWW}}, pages 2789--2797, 2022{\natexlab{b}}.

\bibitem[Tao et~al.(2024)Tao, Wang, Liu, Wu, and Wang]{tao24gard}
Xiang Tao, Liang Wang, Qiang Liu, Shu Wu, and Liang Wang.
\newblock Semantic evolvement enhanced graph autoencoder for rumor detection.
\newblock In \emph{{WWW}}, pages 4150--4159, 2024.

\bibitem[Taylor et~al.(2024)Taylor, Jiang, Qin, Gupta, et~al.]{taylor2024misinformation}
Gabrielle Taylor, Wenting Jiang, Xiao Qin, Ashish Gupta, et~al.
\newblock Misinformation detection: A survey of ai techniques and research opportunities.
\newblock \emph{Foundations and Trends{\textregistered} in Information Systems}, 8\penalty0 (2):\penalty0 66--147, 2024.

\bibitem[Thomas(1986)]{thomas1986statements}
Jeffrey~E Thomas.
\newblock Statements of fact, statements of opinion, and the first amendment.
\newblock \emph{Calif. L. Rev.}, 74:\penalty0 1001, 1986.

\bibitem[Tian et~al.(2022)Tian, Zhang, and Lau]{tian2022duck}
Lin Tian, Xiuzhen~Jenny Zhang, and Jey~Han Lau.
\newblock Duck: Rumour detection on social media by modelling user and comment propagation networks.
\newblock In \emph{Proceedings of the 2022 Conference of the North American Chapter of the Association for Computational Linguistics: Human Language Technologies}, pages 4939--4949, 2022.

\bibitem[Vaswani et~al.(2017)Vaswani, Shazeer, Parmar, Uszkoreit, Jones, Gomez, Kaiser, and Polosukhin]{vaswani2017attention}
Ashish Vaswani, Noam Shazeer, Niki Parmar, Jakob Uszkoreit, Llion Jones, Aidan~N Gomez, {\L}ukasz Kaiser, and Illia Polosukhin.
\newblock Attention is all you need.
\newblock \emph{Advances in neural information processing systems}, 30, 2017.

\bibitem[Vieweg et~al.(2010)Vieweg, Hughes, Starbird, and Palen]{vieweg2010microblogging}
Sarah Vieweg, Amanda~L Hughes, Kate Starbird, and Leysia Palen.
\newblock Microblogging during two natural hazards events: what twitter may contribute to situational awareness.
\newblock In \emph{Proceedings of the SIGCHI conference on human factors in computing systems}, pages 1079--1088, 2010.

\bibitem[Vosoughi et~al.(2018)Vosoughi, Roy, and Aral]{vosoughi2018spread}
Soroush Vosoughi, Deb Roy, and Sinan Aral.
\newblock The spread of true and false news online.
\newblock \emph{science}, 359\penalty0 (6380):\penalty0 1146--1151, 2018.

\bibitem[Wang et~al.(2024)Wang, Wang, Wu, and Liu]{wang2024bots}
Lanjun Wang, Zehao Wang, Le~Wu, and An-An Liu.
\newblock Bots shield fake news: Adversarial attack on user engagement based fake news detection.
\newblock In \emph{Proceedings of the 33rd ACM International Conference on Information and Knowledge Management}, pages 2369--2378, 2024.

\bibitem[Wang et~al.(2012)Wang, Jin, Liu, van~de Koppel, and Alonso]{wang2012simple}
Rong-Hua Wang, Zhen Jin, Quan-Xing Liu, Johan van~de Koppel, and David Alonso.
\newblock A simple stochastic model with environmental transmission explains multi-year periodicity in outbreaks of avian flu.
\newblock \emph{PloS one}, 7\penalty0 (2):\penalty0 e28873, 2012.

\bibitem[Wei et~al.(2021)Wei, Hu, Zhou, Yue, and Hu]{wei2021ebgcn}
Lingwei Wei, Dou Hu, Wei Zhou, Zhaojuan Yue, and Songlin Hu.
\newblock Towards propagation uncertainty: Edge-enhanced bayesian graph convolutional networks for rumor detection.
\newblock In \emph{{ACL/IJCNLP} {(1)}}, pages 3845--3854, 2021.

\bibitem[Wei et~al.(2024)Wei, Hu, Zhou, Wang, and Hu]{wei2022fgcn}
Lingwei Wei, Dou Hu, Wei Zhou, Xin Wang, and Songlin Hu.
\newblock Modeling the uncertainty of information propagation for rumor detection: {A} neuro-fuzzy approach.
\newblock \emph{{IEEE} Trans. Neural Networks Learn. Syst.}, 35\penalty0 (2):\penalty0 2522--2533, 2024.

\bibitem[Wu and Hooi(2023)]{wu2023decor}
Jiaying Wu and Bryan Hooi.
\newblock Decor: Degree-corrected social graph refinement for fake news detection.
\newblock In \emph{SIGKDD}, pages 2582--2593, 2023.

\bibitem[Wu et~al.(2024)Wu, Lin, Cao, and Lin]{wu2024interpretable}
Kaixuan Wu, Yanghao Lin, Donglin Cao, and Dazhen Lin.
\newblock Interpretable short video rumor detection based on modality tampering.
\newblock In \emph{LREC-COLING 2024}, pages 9180--9189. ELRA and ICCL, 2024.

\bibitem[Xu et~al.(2025)Xu, Gao, Li, and Wang]{xu2025d2}
Haowei Xu, Chao Gao, Xianghua Li, and Zhen Wang.
\newblock D2: Customizing two-stage graph neural networks for early rumor detection through cascade diffusion prediction.
\newblock In \emph{Proceedings of the Eighteenth ACM International Conference on Web Search and Data Mining}, pages 568--576, 2025.

\bibitem[Xu et~al.(2018)Xu, Hu, Leskovec, and Jegelka]{xu2018powerful}
Keyulu Xu, Weihua Hu, Jure Leskovec, and Stefanie Jegelka.
\newblock How powerful are graph neural networks?
\newblock \emph{arXiv preprint arXiv:1810.00826}, 2018.

\bibitem[Yang et~al.(2022)Yang, Wang, Jin, Li, Lian, and Xie]{yang2022reinforcement}
Ruichao Yang, Xiting Wang, Yiqiao Jin, Chaozhuo Li, Jianxun Lian, and Xing Xie.
\newblock Reinforcement subgraph reasoning for fake news detection.
\newblock In \emph{Proceedings of the 28th ACM SIGKDD Conference on Knowledge Discovery and Data Mining}, pages 2253--2262, 2022.

\bibitem[Yang et~al.(2024)Yang, Gao, Ma, Lin, and Wang]{yang2024reinforcement}
Ruichao Yang, Wei Gao, Jing Ma, Hongzhan Lin, and Bo~Wang.
\newblock Reinforcement tuning for detecting stances and debunking rumors jointly with large language models.
\newblock \emph{arXiv preprint arXiv:2406.02143}, 2024.

\bibitem[Yang et~al.(2021)Yang, Lyu, Tian, Liu, Liu, and Zhang]{yang2021rumor}
Xiaoyu Yang, Yuefei Lyu, Tian Tian, Yifei Liu, Yudong Liu, and Xi~Zhang.
\newblock Rumor detection on social media with graph structured adversarial learning.
\newblock In \emph{Proceedings of the twenty-ninth international conference on international joint conferences on artificial intelligence}, pages 1417--1423, 2021.

\bibitem[Yang et~al.(2025)Yang, Wang, Zhang, Wang, Wang, and Lam]{yang2025macro}
Xuankai Yang, Yan Wang, Xiuzhen Zhang, Shoujin Wang, Huaxiong Wang, and Kwok~Yan Lam.
\newblock A macro-and micro-hierarchical transfer learning framework for cross-domain fake news detection.
\newblock In \emph{Proceedings of the ACM on Web Conference 2025}, pages 5297--5307, 2025.

\bibitem[Yang et~al.(2023)Yang, Zhao, Wang, Min, Wang, and Wang]{yang2023multimodal}
Yuxing Yang, Junhao Zhao, Siyi Wang, Xiangyu Min, Pengchao Wang, and Haizhou Wang.
\newblock Multimodal short video rumor detection system based on contrastive learning.
\newblock \emph{arXiv preprint arXiv:2304.08401}, 2023.

\bibitem[Yuan et~al.(2019)Yuan, Ma, Zhou, Han, and Hu]{yuan2019jointly}
Chunyuan Yuan, Qianwen Ma, Wei Zhou, Jizhong Han, and Songlin Hu.
\newblock Jointly embedding the local and global relations of heterogeneous graph for rumor detection.
\newblock In \emph{2019 IEEE international conference on data mining (ICDM)}, pages 796--805. IEEE, 2019.

\bibitem[Yuan et~al.(2020)Yuan, Ma, Zhou, Han, and Hu]{yuan2020early}
Chunyuan Yuan, Qianwen Ma, Wei Zhou, Jizhong Han, and Songlin Hu.
\newblock Early detection of fake news by utilizing the credibility of news, publishers, and users based on weakly supervised learning.
\newblock \emph{arXiv preprint arXiv:2012.04233}, 2020.

\bibitem[Zeng et~al.(2025)Zeng, Ding, Cai, Liu, and Qin]{zeng2025exploring}
Yirong Zeng, Xiao Ding, Bibo Cai, Ting Liu, and Bing Qin.
\newblock Exploring large language models for effective rumor detection on social media.
\newblock In \emph{Proceedings of the 2025 Conference of the Nations of the Americas Chapter of the Association for Computational Linguistics: Human Language Technologies (Volume 1: Long Papers)}, pages 2537--2552, 2025.

\bibitem[Zhang et~al.(2022)Zhang, Liang, Yu, and Zhang]{zhang2022rdlt}
Guixian Zhang, Rongjiao Liang, Zhongyi Yu, and Shichao Zhang.
\newblock Rumour detection on social media with long-tail strategy.
\newblock In \emph{2022 International Joint Conference on Neural Networks (IJCNN)}, pages 1--8. IEEE, 2022.

\bibitem[Zhang et~al.(2024{\natexlab{a}})Zhang, Liu, Yang, Yang, Qi, Qian, and Xu]{zhang2024t3rd}
Huaiwen Zhang, Xinxin Liu, Qing Yang, Yang Yang, Fan Qi, Shengsheng Qian, and Changsheng Xu.
\newblock T3rd: Test-time training for rumor detection on social media.
\newblock In \emph{Proceedings of the ACM Web Conference 2024}, pages 2407--2416, 2024{\natexlab{a}}.

\bibitem[Zhang et~al.(2024{\natexlab{b}})Zhang, Li, Liu, Wu, Wang, and Wang]{zhang2024fade}
Jiajun Zhang, Zhixun Li, Qiang Liu, Shu Wu, Zilei Wang, and Liang Wang.
\newblock Evolving to the future: Unseen event adaptive fake news detection on social media.
\newblock In \emph{Proceedings of the 33rd ACM International Conference on Information and Knowledge Management}, pages 4273--4277, 2024{\natexlab{b}}.

\bibitem[Zhang et~al.(2023)Zhang, Yu, Shi, Liang, and Zhang]{zhang23dcerd}
Kaiwei Zhang, Junchi Yu, Haichao Shi, Jian Liang, and Xiaoyu Zhang.
\newblock Rumor detection with diverse counterfactual evidence.
\newblock In \emph{{SIGKDD}}, pages 3321--3331, 2023.

\bibitem[Zhang et~al.(2025)Zhang, Xie, Zhang, Dong, and Wang]{dataset_weibo22}
Yusong Zhang, Kun Xie, Xingyi Zhang, Xiangyu Dong, and Sibo Wang.
\newblock Rumor detection on social media with reinforcement learning-based key propagation graph generator.
\newblock In \emph{Proceedings of the ACM on Web Conference 2025}, pages 2742--2753, 2025.

\bibitem[Zhao et~al.(2024)Zhao, Zhang, Ye, Lu, Yin, and Wang]{zhao2024adversarial}
Mengnan Zhao, Lihe Zhang, Jingwen Ye, Huchuan Lu, Baocai Yin, and Xinchao Wang.
\newblock Adversarial training: A survey.
\newblock \emph{arXiv preprint arXiv:2410.15042}, 2024.

\bibitem[Zhu et~al.(2024)Zhu, Gao, Yin, Li, and Kurths]{zhu2024propagation}
Junyou Zhu, Chao Gao, Ze~Yin, Xianghua Li, and J{\"u}rgen Kurths.
\newblock Propagation structure-aware graph transformer for robust and interpretable fake news detection.
\newblock In \emph{Proceedings of the 30th ACM SIGKDD Conference on Knowledge Discovery and Data Mining}, pages 4652--4663, 2024.

\bibitem[Zubiaga et~al.(2016{\natexlab{a}})Zubiaga, Liakata, and Procter]{dataset_pheme5_1}
Arkaitz Zubiaga, Maria Liakata, and Rob Procter.
\newblock Learning reporting dynamics during breaking news for rumour detection in social media. arxiv 2016.
\newblock \emph{arXiv preprint arXiv:1610.07363}, 2016{\natexlab{a}}.

\bibitem[Zubiaga et~al.(2016{\natexlab{b}})Zubiaga, Liakata, Procter, Wong Sak~Hoi, and Tolmie]{dataset_pheme9}
Arkaitz Zubiaga, Maria Liakata, Rob Procter, Geraldine Wong Sak~Hoi, and Peter Tolmie.
\newblock Analysing how people orient to and spread rumours in social media by looking at conversational threads.
\newblock \emph{PloS one}, 11\penalty0 (3):\penalty0 e0150989, 2016{\natexlab{b}}.

\bibitem[Zubiaga et~al.(2017)Zubiaga, Liakata, and Procter]{dataset_pheme5_2}
Arkaitz Zubiaga, Maria Liakata, and Rob Procter.
\newblock Exploiting context for rumour detection in social media.
\newblock In \emph{Social Informatics: 9th International Conference, SocInfo 2017, Oxford, UK, September 13-15, 2017, Proceedings, Part I 9}, pages 109--123. Springer, 2017.

\bibitem[Zubiaga et~al.(2018)Zubiaga, Aker, Bontcheva, Liakata, and Procter]{zubiaga2018detection}
Arkaitz Zubiaga, Ahmet Aker, Kalina Bontcheva, Maria Liakata, and Rob Procter.
\newblock Detection and resolution of rumours in social media: A survey.
\newblock \emph{Acm Computing Surveys (Csur)}, 51\penalty0 (2):\penalty0 1--36, 2018.

\end{thebibliography}

\end{document}